\documentclass{article}
\usepackage[dvips]{graphicx}
\usepackage{verbatim}
\usepackage{amssymb}
\usepackage{amsmath}
\usepackage{amsbsy}
\usepackage{amsfonts}
\usepackage{amsxtra}
\usepackage{t1enc}
\usepackage{bbm}
\usepackage{mathrsfs}
\usepackage{ifthen}
\usepackage{subfigure}
\def\authors#1{%
  \noindent%
  #1\par%
  }
\def\institute#1{%
  \noindent%
  #1
  }
\def\email#1{%
  \noindent%
  email:
  {\small \texttt #1}
  }
\def\abstract#1{%
  \ifhmode\par\fi
  \removelastskip
  \vskip 5ex\goodbreak
  \noindent
  \leavevmode
  \begingroup
  {\bfseries
  Abstract:} \quad
  #1
  \endgroup%
  \noindent%
  }%
\makeatletter
\renewcommand\section{\@startsection {section}{1}{\z@}%
                                   {-2.5ex \@plus -1ex \@minus -.2ex}%
                                   {.8ex \@plus.2ex}%
                                   {\normalfont\large\bfseries}}
\input{epsf}

\newtheorem{thm}{Theorem}

\newtheorem{cor}{Corollary}

\newcommand{\vphi} {\varphi}
\newcommand{\eps} {\varepsilon}

\newcommand{\bH} {\mathbb{H}}  
\newcommand{\bC} {\mathbb{C}}  
\newcommand{\bR} {\mathbb{R}}  
\newcommand{\bN} {\mathbb{N}}  
\newcommand{\bQ} {\mathbb{Q}}  
\newcommand{\sD} {\mathcal{D}} %
\newcommand{\sL} {\mathcal{L}} %
\newcommand{\sP} {\mathcal{P}} %
\newcommand{\sF} {\mathcal{F}} %
\newcommand{\sM} {\mathcal{M}} %
\newcommand{\sU} {\mathcal{U}} %

\newcommand{\PR} {\mathbb{P}} 
\newcommand{\EX} {\mathbb{E}}

\newcommand{\ud} {\mathrm{d}}

\newcommand{\pder}[1] {\frac{\partial}{\partial #1}}
\newcommand{\ppder}[1] {\frac{\partial^2}{\partial #1^2}}

\newcommand{\til} {\Tilde}

\newcommand{\Span} {\mathrm{span } \; }

\newcommand{\Res}[1] {\mathrm{Res}_{#1} \; }

\newcommand{\ket} {\big>}
\newcommand{\im} {\Im \textrm{m }}

\newcommand{\cl} {\overline}
\newcommand{\bdr} {\partial}

\newcommand{\vir} {\mathfrak{vir}}

\begin{document}

\title{SLE Local Martingales, Reversibility and Duality}
\author{}
\date{}
\maketitle
\authors{Kalle Kytölä and Antti Kemppainen}
\institute{Department of Mathematics and Statistics\\
    P.O.Box 68, FIN-00014 University of Helsinki, Finland.\\
    \email{kalle.kytola@helsinki.fi and antti.h.kemppainen@helsinki.fi}}

%


\abstract{%
We study SLE reversibility and duality using the Virasoro
structure of the space of local martingales. For both problems
we formulate a setup where the questions boil down to comparing two
processes at a stopping time. We state algebraic results
showing that local martingales for the processes have enough in
common. When one has in addition integrability, the method
gives reversibility and duality for any polynomial expected
value.}

\section{Introduction}%
Schramm-Loewner Evolutions (SLEs) are conformally invariant random
curves in two dimensions and their most important properties are
determined by one parameter $\kappa \geq 0$.
SLEs provide insight and a powerful method
to global geometric questions in conformally invariant 2-d
statistical physics at criticality.
Therefore they complement the conformal field theory methods.
SLEs have been successful in obtaining rigorous results about
continuum limits of critical percolation ($\kappa=6$)
\cite{Smirnov:percolation},
loop erased random walk ($\kappa=2$),
uniform spanning tree ($\kappa=8$) \cite{LSW:lerw and ust}
and massless free field level lines ($\kappa=4$) \cite{SS:free field}
but one expects results for many other models as well.
In addition to the question of applying SLE to specific models of
statistical physics, one
can ask questions about SLEs themselves.
In the seminal article \cite{RS:basic properties} many fundamental
properties of SLEs were worked out.
Among the most important open problems,
that paper states conjectures of reversibility and duality.

A chordal SLE is a random curve connecting two points on the boundary.
The clever method of Loewner makes the whole SLE industry possible,
but at the same time the description is
made asymmetric by declaring one a starting point and the other an
end point.
SLE is said to be reversible if the curve is the same when we change
the roles of the two points. Almost without an exception the question
of reversibility is
immediate in models of statistical mechanics. In fact, reversibility
is known for SLE for some values of the parameter $\kappa$ because
of work that
shows that SLE${}_\kappa$ is the continuum limit of some model.
Hidden in our approach to reversibility are
conformal field theory concepts that again bring the starting and end points
to the same status: the operators at the two points have the same
conformal weight $h(\kappa)$ and both have a vanishing
descendant at level two. The vanishing descendants manifest
themselves in our formalism as null field equations for a partition
function $Z$.

If the reversibility property is obvious in models of statistical
mechanics, one might think that SLE reversibility is not a particularly
interesting question from physics point of view.
But conversely, failure of SLE reversibility would mean
losing hope of describing the continuum limit of physical models by SLEs.



Duality is a conjectural property of SLEs that is likely to give a
new kind of geometric insight to two dimensional critical phenomena.
The conjecture relates SLEs with two parameter values where the SLEs
have totally different behavior.
The statement of the conjecture was originally vague:
for $\kappa<4$ the boundary
of SLE${}_{16/\kappa}$ hull looks locally like the SLE${}_\kappa$
trace. This conjecture is supported by considerations of fractal
dimensions, a few examples of models of statistical mechanics and
yet another conformal field theory concept: the central charge $c$,
which takes the same value for SLE${}_\kappa$ and SLE${}_{16/\kappa}$,
i.e. $c(\kappa) = c(16/\kappa)$.
We don't claim to provide a satisfactory
explanation of the origin of duality, but working on the
precise form of the conjecture by Dub\'edat
\cite{D:duality,D:commuting SLEs}, we show an algebraic
reason for a class of expected values to possess the duality property.

As opposed to reversibility, duality seems directly physically relevant.
As an example, it is believed that in critical $q$ state Potts model
for $q \leq 4$, spin cluster boundaries in the continuum limit should be
SLE${}_{\kappa(q)}$ curves with
$\kappa(q) = 4 \cos^{-1} (-\sqrt{q}/2) / \pi \leq 4$.
Potts models admit also a Fortuin-Kasteleyn random cluster
model description. The boundaries of these FK clusters should look
like SLE${}_{16/\kappa(q)}$ 
for $q \in [0,4]$.
Duality would relate these different physical objects in a 
nontrivial geometric way.
Besides the Potts model, there might be other cases of similar type.
For $O(n)$ model in its graphical expansion, spin-spin correlation
functions involve lattice curves connecting the points of insertion of
spins. At critical point and as lattice mesh goes to zero,
these curves for $n \in [0,2]$ are
conjectured to become SLE${}_{\kappa(n)}$,
where $\kappa(n) = 4 \pi / \cos^{-1} (-n/2) \in [8/3 , 4]$.
Since $O(n)$ model allows rewritings of the same kind as the Potts model
\cite{Nienhuis:critical behavior}, it would be interesting to know
whether these involve objects whose scaling limit is
SLE${}_{16/\kappa(n)}$ and whether SLE duality gives insights regarding
this. The relation to SLE of many statistical mechanics models is
reviewed in \cite{Kager:2004yd}.

This letter introduces a setup for the questions of reversibility
and duality using the Virasoro module structure of the space of
local martingales explained in \cite{Kytola:local mgales}.
We state algebraic results supporting both conjectures.
The aim is to compute the behavior of martingales as
distances between certain points tend to zero. Underlying the computations
must be the CFT concepts of fusion and operator product expansions.
In forthcoming articles \cite{KKM:in preparation}
we will provide more careful proofs, discuss the mathematics in
more detail and apply a wider set of methods.

\section{Schramm-Loewner Evolutions}\label{sec: SLE}%
The definition of SLEs appropriate for this note is most
conveniently given in the half plane
$\bH = \{ z \in \bC : \im z > 0 \}$ and allowing the level of
generality of \cite{Kytola:2005kb}.
For comprehensive introduction to SLE we recommend e.g.
\cite{W:lectures, Kager:2004yd, Bauer:2006gr}.
The SLE map $g_t$ is a solution of the Loewner equation
\begin{align*}
\frac{\ud}{\ud t} g_t(z) = \frac{2}{g_t(z) - X_t}
\end{align*}
with initial condition $g_0(z)=z$. 
The map $g_t$ is conformal from $\bH \setminus K_t$ to $\bH$,
where $K_t$ is called the SLE hull at time $t$.
The Loewner equation involves a real valued process
$t \mapsto X_t \in \bR = \bdr \bH$, the driving process,
and we also allow dependency on a number of real points
$Y^K_t=g_t(Y^K_0)$, $K=1,\ldots,M$, that follow passively
the Loewner flow. We assume the driving process to solve the
It\^o stochastic differential equation
\begin{align*}
\ud X_t = \sqrt{\kappa} \; \ud B_t
    + \kappa (\pder{x} \log Z) (X_t ; Y^1_t, \ldots, Y^M_t) \; \ud t
\textrm{ ,}
\end{align*}
where $Z(x; y_1, \ldots, y_M)$ is the partition function
(auxiliary function), which is annihilated by the operator
\begin{align*}
\sD^{(x)} = \frac{\kappa}{2} \frac{\partial^2}{\partial x^2}
    + \sum_{K=1}^M \big( \frac{2}{y_K - x} \pder{y_K}
    - \frac{2 \delta_{y_K}}{(y_K-x)^2} \big)
\textrm{ .}
\end{align*}
The numbers $\delta_{y_K}$ are called the conformal weights at
points $Y^K_0$. The equation $\sD^{(x)} Z = 0$ is called a
null field equation and it is interpreted in conformal field theory
as a vanishing descendant of the operator at the position $x$ of
the driving process.

When $Z$ is of a product form, the process is
SLE${}_\kappa(\rho_1, \ldots, \rho_M)$, introduced in \cite{D:duality}
in the course of studying SLE duality.
The concrete expression
\begin{align*}
Z(x; y_1, \ldots, y_M) = \Big( \prod_{K=1}^M
    (y_K-x)^{\rho_K/\kappa} \Big) \Big( \prod_{1 \leq J < K \leq M}
    (y_J - y_K)^{\rho_J \rho_K / 2 \kappa} \Big)
\end{align*}
was given in \cite{Kytola:2005kb} and the conformal weights are
$\delta_{y_K} = \rho_K (\rho_K + 4 - \kappa) / 4 \kappa$.

The SLE trace $\gamma$ is the (random) curve in $\cl{\bH}$ defined by
$\gamma(t) = \lim_{\eps \downarrow 0} g^{-1}_t (X_t+i\eps)$.
Existence of the limit and continuity of $t \mapsto \gamma(t)$ were
proved in \cite{RS:basic properties}.
The hull is generated by the trace in the sense that $\bH \setminus K_t$
is the unbounded component of $\bH \setminus \gamma[0,t]$.
There is a transition in the qualitative properties of the trace:
for $\kappa \leq 4$ the trace $\gamma$ is a simple path and
$K_t = \gamma[0,t]$ whereas for $\kappa > 4$ we have
$\gamma[0,t] \subsetneq K_t$ and the trace touches itself
and $\bdr \bH = \bR$.

\section{The Virasoro module $\sM$ of local martingales}%
In \cite{Kytola:local mgales} one of us showed how the local
martingales of SLE form a Virasoro module. We briefly explain
the result.

Denote the formal
power series in $z$ whose coefficients are the
independent variables $f_m$, $m \leq -2$, by
\begin{align*}
f(z) = z + \sum_{m \leq -2} f_m z^{1+m}
\textrm{ .}
\end{align*}
Notations such as $f'(z)$ and $Sf(z) =
\frac{f'''(z)}{f'(z)} - \frac{3}{2} \frac{f''(z)^2}{f'(z)^2}$
and rational functions of $f$ are understood as formal power
series at infinity, always containing only finitely many
positive powers of the argument.
Residue of a formal power series in $z$ is the coefficient
of $z^{-1}$. 
Given $\delta_{(\cdot)}$, $Z$ as in Section \ref{sec: SLE} and
$c \in \bC$, we can define the operators
\begin{align*}
\sL_n  \, = \, & \Res{r} r^{1-n} \Big\{ \frac{c}{12} Sf(r)
    + \frac{\delta_x \, f'(r)^2}{(f(r)-x)^2}
    + \frac{\partial_x Z}{Z} \; \frac{f'(r)^2}{f(r)-x} \\
& \qquad + \sum_K \big( \frac{\delta_{y_K} \, f'(r)^2}{(f(r)-y_K)^2}
    + \frac{\partial_{y_K} Z}{Z} \; \frac{f'(r)^2}{f(r)-y_K} \big) \\
& \qquad + \frac{f'(r)^2}{f(r)-x} \; \pder{x}
    + \sum_K \frac{f'(r)^2}{f(r)-y_K} \; \pder{y_K} \\
& \qquad - \sum_{l \leq -2} \Res{z} z^{-2-l} \frac{f'(r)^2}{f(z)-f(r)}
    \; \pder{f_l} \Big\}
\end{align*}
on a suitable function space
$\sF \big( x, (y_K)_{1 \leq K \leq M}, (f_l)_{l \leq -2} \big)$.
A mere change of notation from \cite{Kytola:local mgales}
shows that the operators $\sL_n$
satisfy the commutation relations of the Virasoro algebra $\vir$
(for Virasoro algebra and its representations see e.g. \cite{Kac}
and \cite{FF:long paper}).
For both geometric and algebraic reasons we assign degree $1$ to
the variables $x, y_1, \ldots, y_M$ and degree $m$ to $f_{-m}$.
The degree of a monomial is the sum of degrees of its variables,
counting multiplicities.

Local martingales are functions $\vphi$ such that the It\^o
derivative of \linebreak[4]$\vphi(X_t, \ldots, Y^M_t ; g_{-2}(t), \ldots)$
has no drift term, i.e.
\begin{align*}
\ud \vphi(X_t ; Y^1_t, \ldots, Y^M_t ; g_{-2}(t), g_{-3}(t), \ldots)
    =  0 \; \ud t + (\cdots) \; \ud B_t
\textrm{ .}
\end{align*}
The operators $\sL_n$ were shown to preserve the space of local martingales
for the specific values
$\delta_x=h(\kappa)=\frac{6-\kappa}{2 \kappa}$ and
$c=c(\kappa) = \frac{(3-8\kappa)(6-\kappa)}{2 \kappa}$.
Starting from the constant function $1$ and applying in all
possible ways the operators $\sL_n$ one generates the $\vir$
module
\begin{align*}
\sM = \sU(\vir) \cdot 1
\end{align*}
that consists of local martingales.
In fact, as shown in \cite{Kytola:local mgales}, if $Z$ is translation
invariant and homogeneous, $\sM$ is a highest weight module for $\vir$
with the constant function $1$ as its highest weight vector.

For $\kappa \notin \bQ$, any Verma module for $\vir$ with
central charge $c(\kappa)$ is either irreducible or contains
a maximal submodule generated by a single singular vector
\cite{FF:long paper}.
We refer to this case as generic $\kappa$.

\section{Setup for Reversibility}\label{sec: reversibility}%
The chordal SLE from $0$ to $\infty$ in $\bH$ can be viewed as an
SLE with no extra points $Y$ and constant partition function $Z(x)=1$.
The reversibility conjecture states that the trace $\gamma$
has the same law as the image of $\gamma$ under the inversion
$z \mapsto -1/z$ of $\bH$.
The latter is the trace of an SLE from $\infty$ to $0$ in $\bH$.

\begin{figure}
\includegraphics[width=1.0\textwidth]{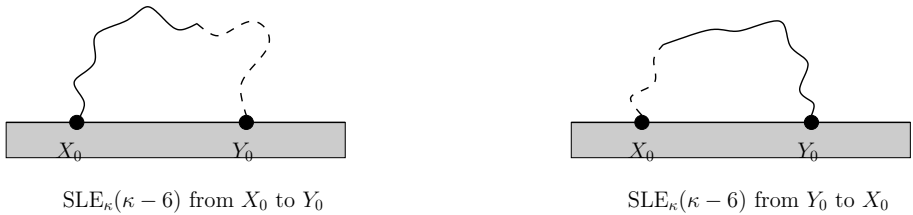}
\caption{The two processes for reversibility.}
\label{fig: reversibility}
\end{figure}

For the question of reversibility we find it more convenient to compare
the chordal SLE from $X_0 \in \bR$ to $Y_0 \in \bR$ and
from $Y_0$ to $X_0$, see Figure \ref{fig: reversibility}.
This is obtained by a M\"obius coordinate change from the usual
chordal SLE, see e.g. \cite{SW, Bauer:2005jt}.
This variant is SLE${}_\kappa (\rho)$, $\rho = \kappa-6$, and the
appropriate partition function is
$Z(x,y)=(y-x)^{\frac{\kappa - 6}{\kappa}}$.
The conformal weight at the driving process and at the endpoint is
the same, $\delta_x = \delta_y = h(\kappa) = \frac{6-\kappa}{2 \kappa}$.
The Loewner equation for an SLE from $X_0$ to $Y_0$ is
$\dot{g}^+_t = 2/(g^+_t - X^+_t)$, where the driving process is
$\ud X^+_t = \sqrt{\kappa} \; \ud B_t +
\kappa (\partial_{x} \log Z) \; \ud t$ and the other point
$Y^+_t = g^+_t(Y_0)$ is passive.
The process is defined up to the stopping time $\tau^+$ at which
$\lim_{t \uparrow \tau^+} | Y^+_t - X^+_t | = 0$. At the end,
$g^+_{\tau^+}$ maps the ``outside'' $\bH \setminus K^+_{\tau^+}$ of the
SLE trace conformally onto $\bH$.
To get a physical picture, consider for example the Ising model
(believed to correspond to $\kappa=3$).
Imposing boundary conditions $\uparrow$ on $[X_0,Y_0]$ and $\downarrow$
on the rest of the real axis, the hull $K^+_{\tau^+}$ would be a
component disconnected from $\infty$ by the curve $\gamma$
from $X_0$ to $Y_0$
that follows spin cluster boundaries.

The reverse case, an SLE from $Y_0$ to $X_0$ is obtained with the
same partition function --- one should observe that
$Z(x,y)=(y-x)^{\frac{\kappa-6}{\kappa}}$ is annihilated not only
by $\sD^{(x)}$ but also by
\begin{align*}
\sD^{(y)} = \frac{\kappa}{2}\ppder{y} + \frac{2}{x-y} \pder{x}
    - \frac{2 \delta_x}{(x-y)^2}
\textrm{ .}
\end{align*}
The Loewner equation $\dot{g}^-_t = 2/(g^-_t - Y^-_t)$ has
$Y^-_t$ as its driving process,
$\ud Y^-_t = \sqrt{\kappa} \; \ud B_t +
\kappa (\partial_{y} \log Z) \; \ud t$, and as a passive point
$X^-_t = g^-_t(X_0)$.
At stopping time $\tau^-$ when $Y^-$ and $X^-$ collide,
$K^-_{\tau^-}$ and $g^-_{\tau^-}$ are expected to have
the same law as $K^+_{\tau^+}$ and $g^+_{\tau^+}$ in the
non-reversed case --- this is precisely the content of the
reversibility conjecture for $\kappa \leq 4$.

Before we start the general consideration, let's give a concrete
illustration of the technique. The coefficient $g_{-2}(t)$,
called the half-plane capacity, measures the size of $K_t$:
for example if $g_{-2}(t) \leq R^2$ then the radius of $K_t$ is not
more than $R$.
The function $\sL_{-2} \cdot 1 \in \sM$ is easily computed to be
$- f_{-2} \, c(\kappa)/2 + (y-x)^2 \, h(\kappa)$. 
Therefore $-\frac{c(\kappa)}{2} g^+_{-2}(t) + h(\kappa) \, (Y^+_t-X^+_t)^2$
is a local martingale. Supposing that it is in fact a closable martingale
(if $\EX [g_{-2}^+(\tau^+)] < \infty$, it is), we can compute
the average of $g_{-2}^+(\tau^+)$ because expected values of
martingales are constant in time
\begin{align*}
\EX [ -\frac{c(\kappa)}{2} g_{-2}^+(\tau^+) ]
\; = \; & \EX \big[ \big(\sL_{-2} \cdot 1\big) ( X^+_{\tau^+},Y^+_{\tau^+};
    g_{-2}^+(\tau^+) ) \big]  \\
\; = \; & \EX \big[ \big(\sL_{-2} \cdot 1\big)
    ( X^+_0,Y^+_0; g_{-2}^+(0) ) \big] 
= \EX [h(\kappa) (Y_0-X_0)^2]
\textrm{ .}
\end{align*}
Here we read that the average size of $K^+_{\tau^+}$ in terms of capacity
is $\frac{2}{8-3\kappa} (Y_0-X_0)^2$, which makes sense for $\kappa < 8/3$.
The same can be done with
$g_{-2}^-(\tau^-)$ and one finds that at least the average capacity is
same for the reversed case (this is not new, though). Our strategy is to
pick more complicated local martingales from $\sM$ to determine more
general expected values.

Since $Z(x,y)$ is the same for the SLE from $X_0$ to $Y_0$ and the
reversed SLE, the representation $\sM$ is obviously the same for both
cases. Both cases are SLEs in the sense of Section \ref{sec: SLE},
one with driving process $x$ and null field equation $\sD^{(x)} Z = 0$,
the other with driving process $y$ and null field equation
$\sD^{(y)} Z = 0$.
Now as a consequence of $c=c(\kappa)$ and $\delta_x=\delta_y=h(\kappa)$,
for any $\vphi \in \sM = \sU(\vir) \cdot 1$ the process
\begin{align*}
\vphi(X^\pm_t , Y^\pm_t ; g^\pm_{-2}(t), g^\pm_{-3}(t), \ldots)
\end{align*}
is a local martingale for both ``$+$'' and ``$-$''.

The elements $\sL_{-n_1} \cdots \sL_{-n_k} \cdot 1$ that span the
representation $\sM$ are homogeneous polynomials of degree $\sum_j n_j$ in
$x,y,f_{-2},f_{-3},\ldots$ (recall that $f_{-m}$ was assigned a degree $m$).
This is because
$\sL_{-n}$ are differential operators containing only polynomial
multiplications and they raise the degree by $n$.
So $\sM$ is a subspace of the space of polynomials,
$\sM \subset \bC [x,y,f_{-2},\ldots]$.
One also directly checks that $\sL_0 \cdot 1 = 0$ so $\sM$ is a highest
weight representation of highest weight $0$.
By induction, keeping track of
contributions of different parts of the operator $\sL_{n}$
one establishes the important fact that
$\sL_{-n_1} \cdots \sL_{-n_k} \cdot 1$ can be written as
\begin{align*}
P_{n_1, \ldots, n_k}(f_{-2}, \ldots) +
    (y-x) R_{n_1, \ldots, n_k}(x,y;f_{-2},\ldots)
\textrm{ ,}
\end{align*}
where $P$ and $R$ are polynomials.
This captures the behavior of the local martingale as $X$ and $Y$
processes come together, namely only the $P$ part remains in the limit
$|x-y| \rightarrow 0$. The $P$ themselves form a highest weight module
$\sP = \Span \{ P_{n_1, \ldots, n_k} \} \subset
\bC[f_{-2}, f_{-3}, \ldots]$ with highest
weight vector $1$ in the obvious way.

The usefulness of the above is the consequence that local martingales
for both processes have the same initial and final values and dependence
of the quite different stochastic processes $X^\pm_t$, $Y^\pm_t$
disappears in the end.
More precisely, choose $\vphi \in \sM$ and denote its decomposition by
$\vphi = P + (y-x) R$. 
Then $\vphi$ is a local martingale for both the SLE from $X_0$ to $Y_0$
and for the SLE from $Y_0$ to $X_0$. Moreover,
its initial value at $t=0$ is the same for the two processes
\begin{align*}
\vphi(X^\pm_0,Y^\pm_0; g^\pm_{-2}(0), \ldots)
    = P(0,0,\ldots) + (Y_0 - X_0) R(X_0,Y_0; 0,0,\ldots)
\end{align*}
and the final value at $t=\tau^\pm$ is the same function of
the coefficients of $g^\pm_{\tau^\pm}$
\begin{align*}
\vphi(X^\pm_{\tau^\pm},Y^\pm_{\tau^\pm}; g^\pm_{-2}(\tau^\pm), \ldots)
    = P(g^\pm_{-2}(\tau^\pm), g^\pm_{-3}(\tau^\pm), \ldots)
\textrm{ .}
\end{align*}

In the case of reversibility, we can actually make an estimate of
$L^1(\PR)$ norm to show that for $\kappa < 8/(1+\sum_j n_j)$,
$\sL_{-n_1} \cdots \sL_{-n_k} \cdot 1$ is a
closable martingale up to the stopping time $\tau^\pm$.
Using this and the above observations of $\phi \in \sM$, we
establish reversibility of expected values of $P$.
\begin{thm}
Let $\vphi = P +(y-x)R \in \sM$ as above.
For $\kappa$ small enough, the random variables
$P(g^\pm_{-2}(\tau^\pm), g^\pm_{-3} (\tau^\pm), \ldots)$
are integrable, $\vphi(X^\pm_t, Y^\pm_t ; g^\pm_{-2}(t), \ldots)$
are closable martingales up to the stopping times $\tau^\pm$
and consequently
\begin{align*}
\EX [P(g^\pm_{-2}(\tau^\pm), g^\pm_{-3} (\tau^\pm), \ldots)]
    = P(0,0,\ldots) + (Y_0 - X_0) \; R(X_0, Y_0; 0,0,\ldots)
\textrm{ .}
\end{align*}
\end{thm}

Having discussed reversibility we now turn to the other question,
duality. The strategy will be similar, even if the cumbersome
details make it less transparent.

\section{Setup for Duality}%
\begin{figure}
\includegraphics[width=1.0\textwidth]{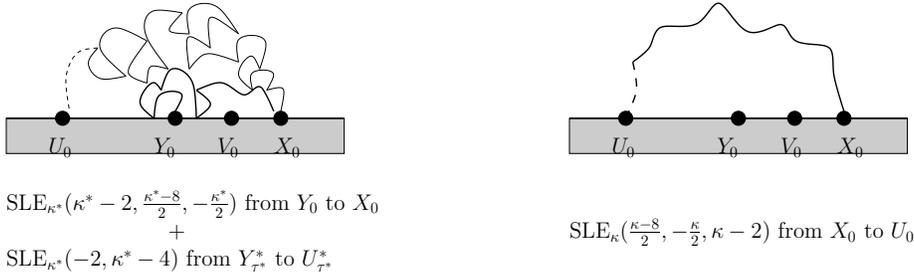}
\caption{The two processes for duality.}
\label{fig: duality}
\end{figure}

Recall that for $\kappa \leq 4$ the SLE${}_\kappa$ trace is a simple
curve. For $\kappa > 4$ the trace generates a strictly larger hull $K_t$,
and the boundary of the hull, $\bdr K_t$, can be parametrized as
a continuous curve.
The duality conjecture states roughly that for $0 < \kappa < 4$
and $\kappa^* = 16/\kappa$, the boundary of the hull of SLE${}_{\kappa^*}$
looks like the trace of SLE${}_\kappa$. The conjecture was formulated
more precisely by Dub\'edat in \cite{D:duality, D:commuting SLEs}.
The processes we consider below are obtained by a coordinate change 
from Dub\'edat's formulation.
The general idea is again to compare the two processes at their stopping
times. The driving process and other points will come together
and we decompose local martingales accordingly. The decompositions
show that we have continuous local martingales with same initial and
final values, exactly as in the case of reversibility.
The setup is explained
in the paragraphs below and illustrated in Figure \ref{fig: duality}.


Fix $\kappa<4$ and points $U_0 < Y_0 < V_0 < X_0$.
Instead of an ordinary SLE${}_\kappa$ we start from $X_0$ an
SLE${}_\kappa (\rho_u , \rho_y , \rho_v)$, where
$\rho_u = \frac{\kappa-8}{2}$, $\rho_y = -\frac{\kappa}{2}$
and $\rho_v = \kappa - 2$.
The partition function is
\begin{align*}
Z \; = \; & (y-u)^{\Delta_{y,u}} (v-u)^{\Delta_{v,u}} (x-u)^{\Delta_{x,u}}
    (v-y)^{\Delta_{v,y}} (x-y)^{\Delta_{x,y}} (x-v)^{\Delta_{x,v}}
\textrm{ ,}
\end{align*}
with the values of $\Delta_{(\cdot,\cdot)}$ and conformal weights
$\delta_{(\cdot)}$ listed in Table \ref{tbl: deltas} (a).
The partition function satisfies $\sD^{(x)} Z = 0$ and the
value of $\delta_x$ is $\delta_x=h(\kappa)$ again.
The Loewner flow is $\dot{g}_t = 2/(g_t - X_t)$ and
$\ud X_t = \sqrt{\kappa} \; \ud B_t +
\kappa (\partial_{x} \log Z) \; \ud t$ whereas the rest of the
points are passive, $U_t = g_t(U_0)$, $Y_t = g_t(Y_0)$,
$V_t = g_t(V_0)$. Such an SLE will start from $X_0$ and end at
$U_0$ at time $\tau$ at which $U_\tau = Y_\tau = V_\tau = X_\tau$.
\begin{table}[tb!]
\centering
\subfigure[]{\small 
\begin{tabular}{@{} r | c @{\quad} c @{\quad} c @{\quad} c | c @{}}
 & $u$&$y$&$v$&$x$&\\
\hline
& & & & &\\[-9pt]
$u$ & &
$\frac{8-\kappa}{8}$ &
$\frac{(\kappa-2)(\kappa-8)}{4 \kappa}$ &
$\frac{\kappa-8}{2\kappa}$ &\\[3pt]

$y$ & & &
$\frac{2-\kappa}{4}$ &
$-\frac{1}{2}$ & $\Delta_{(\cdot, \cdot)}$ \\[3pt]

$v$ & & & &
$\frac{\kappa-2}{\kappa}$ &\\[3pt]
\hline
& & & & &\\[-9pt]
  & 
$\frac{8-\kappa}{16}$& 
$\frac{3\kappa-8}{16}$& 
$\frac{\kappa-2}{2\kappa}$& 
$\frac{6-\kappa}{2\kappa}$& $\delta_{(\cdot)}$
\end{tabular}
}\hspace{0.25cm}
\subfigure[]{\small
\begin{tabular}{@{} r | c @{\quad} c @{\quad} c | c @{}}
 & $\til{u}^*$&$\til{y}^*$&$\til{w}^*$&\\
\hline
& & & &\\[-9pt]
$\til{u}^*$ & &
$- \frac{2}{\kappa^*}$ &
$\frac{4-\kappa^*}{\kappa^*}$  &\\[0pt]
$\til{y}^*$ & & &
$\frac{\kappa^*-4}{\kappa^*}$ & 
\raisebox{6pt}{$\Delta_{(\cdot, \cdot)}$} \\[3pt]
\hline
& & & &\\[-9pt]
  & 
$\frac{\kappa^*-2}{2\kappa^*}$& 
$\frac{6-\kappa^*}{2\kappa^*}$& 
$0$& $\delta_{(\cdot)}$
\end{tabular}
}
\caption{Values of $\Delta$ and $\delta$ in the duality setup}
\label{tbl: deltas}
\end{table}

As in Section \ref{sec: reversibility}, a concrete illustration of the
general technique is determining the average capacity of $K_\tau$.
The appropriate local martingale again comes from
\begin{align*}
\sL_{-2} \cdot 1 = -\frac{c(\kappa)}{2} f_{-2} + u^2 \delta_u + \cdots
    + x^2 \delta_x + uy \, \Delta_{u,y} + \cdots + vx \, \Delta_{v,x},
\textrm{ .}
\end{align*}
Plugging in the processes at times $t=0$ and $t=\tau$ and assuming further
that this gives a closable martingale, one easily reads a (not particularly
enlightening but nevertheless explicit) formula
for the average size of $K_\tau$ in terms of capacity.

We will compare the above variant of SLE${}_\kappa$ to a variant of
SLE with the dual parameter $\kappa^*=16/\kappa$.
This SLE will be glued from two pieces.
First start from $Y_0$ an
SLE${}_{\kappa^*}(\rho^*_{u^*}, \rho^*_{v^*}, \rho^*_{x^*})$, where
$\rho^*_{u^*} = \kappa^*-2$, $\rho^*_{v^*} = \frac{\kappa^*-8}{2}$ and
$\rho^*_{x^*} = -\frac{\kappa^*}{2}$.
The driving process is $Y^*_t$,
$\ud Y^*_t = \sqrt{\kappa^*} \; \ud B_t
+ \kappa^* (\partial_{y^*} Z^*) \; \ud t$, and the rest are passive
$U^*_t = g^*_t(U_0)$, $V^*_t=g^*_t(U_0)$, $X^*_t = g^*_t(X_0)$.
The partition function is the same as above, $Z^* = Z$, and it is
important that it is annihilated also by
\begin{align*}
\sD^{(y^*)} \; = \; & \frac{\kappa^*}{2} \ppder{{y^*}}
    + \frac{2}{u^*-y^*} \pder{u^*} + \frac{2}{v^*-y^*} \pder{v^*}
    + \frac{2}{x^*-y^*} \pder{x^*} \\
& - \frac{2 \delta_{u^*}}{(u^*-y^*)^2} - \frac{2 \delta_{v^*}}{(v^*-y^*)^2}
    - \frac{2 \delta_{x^*}}{(x^*-y^*)^2}
\textrm{ .}
\end{align*}
The conformal weights are the same
(Table \ref{tbl: deltas} (a): $\delta_{u^*}= \delta_u$, \ldots),
but as the driving process is $Y^*_t$,
the value that is important for the local martingales is
$\delta_{y^*} = h(\kappa^*) = \frac{6-\kappa^*}{2\kappa^*}
= \frac{3 \kappa - 8}{16}$ now.
We consider this process up to the first time $\tau^*$
at which the three points $Y^*$, $V^*$ and $X^*$ will collide.
After that we continue from the collision point $Y^*_{\tau^*}$ an
SLE${}_{\kappa^*}(\rho^*_{\til{u}^*},\rho^*_{\til{w}^*})$, where the extra
points are started at $\til{U}^*_{\tau^*} = U^*_{\tau^*}$ and
$\til{W}^*_{\tau^*} = Y^*_{\tau^*} + 0$ with
$\rho^*_{\til{u}^*} = -2$, $\rho^*_{\til{w}^*}=\kappa^*-4$.
This means that we use as the initial value for
$\dot{\til{g}}^*_t = 2 / (\til{g}^*_t - \til{Y}^*_t)$ at $t=\tau^*$
the final value $g^*_{\tau^*}$. Again
$\til{U}^*_t$ and $\til{W}^*_t$ are passive.
The partition function for this part of the process is
\begin{align*}
\til{Z}^* = (\til{y}^*-\til{u}^*)^{\Delta_{\til{y}^*,\til{u}^*}}
    (\til{w}^* - \til{u}^*)^{\Delta_{\til{w}^*,\til{u}^*}}
    (\til{w}^*-\til{y}^*)^{\Delta_{\til{w}^*,\til{y}^*}}
\end{align*}
with $\Delta_{(\cdot,\cdot)}$ and $\delta_{(\cdot)}$ as
in Table \ref{tbl: deltas} (b).
Finally, the driving process $\til{Y}^*_t$ and
$\til{U}^*_t$ will collide at stopping time $\til{\tau}^*$.

For the first SLE it turns out as before that
$\sM = \sU(\vir) \cdot 1 \subset \bC[u,y,v,x, f_{-2}, \ldots]$ is a
highest weight module consisting of local martingales for the
process. The highest weight is $0$ and the module is irreducible
for generic $\kappa$. The $\kappa^*$ SLE was constructed by gluing
two pieces. It will turn out that local martingales are
obtained by gluing, too.

Consider the two representations
$\sM^* \subset \bC[u^*,y^*,v^*,x^*,f_{-2},\ldots]$
and $\til{\sM}^* \subset \bC[\til{u}^*,\til{y}^*,\til{w}^*, f_{-2}, \ldots]$
corresponding to the partition functions $Z^*$ and $\til{Z}^*$.
They, too, are highest weight representations
with highest weight $0$ and irreducible for generic $\kappa$.
We'd like to show that for any $n_1, \ldots, n_k$ the
``glued'' process
\begin{align*}
\big(\sL^*_{n_1} \cdots \sL^*_{n_k} \cdot 1 \big)
(U^*_t, Y^*_t, V^*_t, X^*_t; g^*_{-2}(t), \ldots) 
\quad \textrm{ for $0 \leq t \leq \tau^*$} \\
\big(\til{\sL}^*_{n_1} \cdots \til{\sL}^*_{n_k} \cdot 1 \big)
(\til{U}^*_t, \til{Y}^*_t, \til{W}^*_t; \til{g}^*_{-2}(t), \ldots) 
\quad \textrm{ for $\tau^* < t \leq \til{\tau}^*$}
\end{align*}
is a continuous local martingale for the ``glued'' SLE defined
above. We denote the glued local martingale below by
$\vphi^{\textrm{glued}}$.
The continuity is based on decompositions of the local martingales
in $\sM^*$ and $\til{\sM}^*$. One can write 
$\sL^*_{-n_1} \cdots \sL^*_{-n_k} \cdot 1$ as a
sum of 
$Q_{n_1, \ldots, n_k}^*(u^*,y^*;f_{-2},\ldots)$ and terms that
have factors $(x^*-y^*)$ or $(v^*-y^*)$. 
Similarly,
$\til{\sL}^*_{-n_1} \cdots \til{\sL}^*_{-n_k} \cdot 1$ is a
sum of 
$\til{Q}^*_{n_1, \ldots, n_k}(\til{u}^*,\til{y}^*; f_{-2}, \ldots)$
and terms that have a factor $(\til{w}^*-\til{y}^*)$.
What is needed is the non-trivial fact that $Q^*_{n_1, \ldots, n_k}$
and $\til{Q}^*_{n_1, \ldots, n_k}$ are the same functions.

According to the duality conjecture, the first SLE at time $\tau$
should look the same as the second,
glued SLE, at time $\til{\tau}^*$. So we need to compare the final
values of local martingales in $\sM$ and $\til{\sM}^*$. In order to do
so, we use more decompositions that exhibit the behavior of local
martingales after relevant fusions.
Like for reversibility, induction and splitting $\sL^*_n$ and
$\til{\sL}^*_n$ in parts allow to show that any
$\til{\sL}^*_{-n_1} \cdots \til{\sL}^*_{-n_k} \cdot 1$ can be written as a
sum of $P_{n_1, \ldots, n_k} (f_{-2}, \ldots)$ and
$(\til{y}^*-\til{u}^*) \til{R}^*_{n_1, \ldots, n_k}
(\til{u}^*,\til{y}^*,\til{w}^*;f_{-2}, \ldots)$.
Also, any
$\sL_{-n_1} \cdots \sL_{-n_k} \cdot 1$ can be written as a sum
of $P_{n_1, \ldots, n_k} (f_{-2}, \ldots)$ and
$R_{n_1, \ldots, n_k}(u,y,v,x;f_{-2}, \ldots)$, where $R_{n_1,\ldots,n_k}$
is a sum of terms, each of which has a factor $(y-u)$,
$(v-u)$ or $(x-u)$.
The polynomials $P_{n_1,\ldots, n_k}$ are precisely the ones
occurring also in Section \ref{sec: reversibility}.
Since $Z^*=Z$ we have
$\sL^*_{-n_1} \cdots \sL^*_{-n_k} \cdot 1
= \sL_{-n_1} \cdots \sL_{-n_k} \cdot 1$ so that initial values of
the local martingales are the same.
Again the $P \in \bC[f_{-2}, f_{-3}, \ldots]$ form
the representation $\sP$.

As in the reversibility case, if we have closable martingales,
we can make a conclusion about expected values.
For duality, we can't control in which range of the parameter
$\kappa$ the expected values are finite so the result is less
explicit.
\begin{thm}
Choose $\vphi \in \sM = \sM^*$ and write $\vphi = P + R$ as above.
Then $\vphi$ is a local martingale for the
SLE${}_\kappa(\rho_u,\rho_y,\rho_v)$ and $\vphi^\textrm{glued}$
is a local martingale for the glued SLE${}_{\kappa^*}$.
The initial value for both is $\vphi(U_0,Y_0,V_0,X_0;0,0,\ldots)$ and
the final value is $P$ of the coefficients
\begin{align*}
\vphi |_{t=\tau} = P(g_{-2}(\tau), \ldots)
& \quad \textrm{ and } \; &
\vphi^\textrm{glued} |_{t=\til{\tau}^*}
    = P(\til{g}^*_{-2}(\til{\tau}^*), \ldots)
\textrm{ .}
\end{align*}
If $P(g_{-2}(\tau),\ldots)$ and $P(\til{g}^*_{-2}(\til{\tau}^*),\ldots)$
are integrable, then the local martingales corresponding to $\vphi$
are closable martingales up to times $\tau$ and $\til{\tau}^*$ and
\begin{align*}
\EX [ P(g_{-2}(\tau),\ldots) ] = \EX[ P(\til{g}^*_{-2}(\til{\tau}^*), \ldots) ]
= P(0,\ldots) + R(U_0,Y_0,V_0,X_0;0,\ldots)
\textrm{ ,}
\end{align*}
i.e. duality holds for the expected value of the polynomial $P$.
\end{thm}

\section{Enough local martingales to find all moments}%
So far we have presented setups for reversibility and duality that
allow us to show that for any polynomial
$P(f_{-2}, f_{-3}, \ldots) \in \sP$,
the reversibility and duality hold for those $\kappa$ for which
$P$ at the final stopping time is in $L^1(\PR)$.
The obvious next question is whether $\sP$ contains enough
polynomials for these statements to be useful.
The answer is nice and easy --- for $\kappa$ generic $\sP$
contains \emph{all} polynomials.

Indeed, it is not difficult to show that for $\kappa$ generic, $\sP$ is
the irreducible highest weight representation of highest weight
$0$. This means that there is a null vector
$\sL_{-1} \cdot 1 = 0$. We can write
$\sP = \oplus_{n=0}^\infty \sP^{(n)}$, where $\sP^{(n)}$ is
the (finite dimensional) $\sL_0$ eigenspace of eigenvalue
$n$. It consists of homogeneous polynomials of degree $n$.
For the Verma module, the dimensions of the eigenspaces are
$\dim (\mathrm{Verma}^{(n)}) = p(n) = \# \{ (n_1, \ldots, n_k) : k \in \bN ,
1 \leq n_1 \leq \cdots \leq n_k , n_1 + \cdots + n_k = n\}$.
In the generic case, the Verma module has a maximal submodule
generated by $L_{-1} | 0 \ket$, which itself is a Verma module of
highest weight $1$. The quotient is irreducible and therefore isomorphic
to $\sP$ and we can conclude that the dimensions are
\begin{align*}
\dim (\sP^{(n)}) = p(n) - p(n-1)
\textrm{ .}
\end{align*}

The polynomials $f_{-2}^{m_2} \cdots f_{-l}^{m_l}$ with $l \in \bN$
and $\sum_{j=2}^l j m_j = n$
certainly form a basis for polynomials of degree $n$ in
$f_{-2}, f_{-3}, \ldots$ (remember that $f_{-d}$ is of degree $d$).
The number of these is
$q(n) = \# \{(m_2, \ldots, m_n) \in \bN^{n-1} : \sum_{j=2}^l j m_j = n \}$.
It's easy to check that $q(n) = p(n) - p(n-1) = \dim (\sP^{(n)})$,
which immediately says that
for generic $\kappa$, the space $\sP^{(n)}$ contains all homogeneous
polynomials of degree $n$.
Combining with the $L^1(\PR)$ estimate in reversibility case,
this has the following consequence.
\begin{cor}
\label{cor: conclusion}
Fix $m_2, \ldots, m_l \in \bN$. Then for
$\kappa < 8 / (1 + \sum_{j=2}^l j m_j )$, $\kappa \notin \bQ$
the expected values
\begin{align*}
\EX [ g^\pm_{-2}(\tau^\pm)^{m_2} \cdots g^\pm_{-l}(\tau^\pm)^{m_l} ]
\end{align*}
exist and are equal. Similarly, for
$\kappa \notin \bQ$ such that the the expected values
\begin{align*}
\EX [ g_{-2}(\tau)^{m_2} \cdots g_{-l}(\tau)^{m_l} ]
\quad \textrm{ and } \quad
\EX [ \til{g}^*_{-2}(\til{\tau}^*)^{m_2} \cdots
    \til{g}^*_{-l}(\til{\tau}^*)^{m_l} ]
\end{align*}
exist, they are equal. In other words, reversibility and duality hold
for any monomial expected value, provided it exists.
\end{cor}

\section{Conclusions}%
We have exhibited setups for studying the well known open
problems of reversibility and duality of SLE. An analysis of
the Virasoro module structure of local martingales
leads to statements strongly supporting both conjectures.
For the processes that one has to compare,
we can find enough local martingales of the same functional form
to account for reversibility and duality in an algebraic sense.
However, any given polynomial expected value
only exists up to a certain value of $\kappa$, which is
small when the degree of the polynomial is large.
We will report
on the problems in more detail and using also other methods in
\cite{KKM:in preparation}.

\bigskip

\emph{Acknowledgements.}
We thank Antti Kupiainen and Paolo \linebreak[4] Muratore-Ginanneschi
for discussions and helpful suggestions.
A.K. wants to thank Stanislav Smirnov for discussions
on questions of reversibility and duality. A.K. was
financially supported by Finnish Academy of Science and Letters,
Vilho, Yrjö and Kalle Väisälä Foundation.

\end{document}